\documentclass[pre,aps,showpacs,twocolumn]{revtex4}

\usepackage{epsfig}
\usepackage{graphics}
\usepackage{graphicx}
\usepackage{bm}
\usepackage{color}

\usepackage[normalem]{ulem}

\begin{document}

\title{Visco-rotational shear instability of Keplerian granular flows}

\author{Luka G. Poniatowski}
\email{luka.poniatowski@gmail.com}
\affiliation{Faculty of Exact and Natural Sciences,
Tbilisi State University, 3 Chavchavadze ave., Tbilisi 0179, Georgia}
\affiliation{Abastumani Astrophysical Observatory, Ilia State University, 2
G. Tsereteli str., Tbilisi 0162, Georgia}

\author{Alexander G.\ Tevzadze}
\email{aleko@tevza.org}
\affiliation{Faculty of Exact and Natural Sciences,
Tbilisi State University, 3 Chavchavadze ave., Tbilisi 0179, Georgia}
\affiliation{Abastumani Astrophysical Observatory, Ilia State University, 2
G. Tsereteli str., Tbilisi 0162, Georgia}

\begin{abstract}
The linear stability of viscous Keplerian flow around a gravitating center
is studied using the rheological granular fluid model. The linear
rheological instability triggered by the interplay of the shear rheology
and Keplerian differential rotation of incompressible dense granular fluids
is found. Instability sets in granular fluids, where the viscosity
parameter grows faster than the square of the local shear rate (strain
rate) at constant pressure. Found instability can play a crucial role in
the dynamics of dense planetary rings and granular flows in protoplanetary
disks.
\end{abstract}



\maketitle

Disk of solid particles rotating around central gravitating object is an
important class of granular flows widely occurring in nature. Among those are
planetary and exoplanetary rings, debris disks around young stars, or even
areas of protoplanetary disks where dust particles accumulate and form dense
granular material.
These flows, occurring at different scales, often have several common
features: solid particles rotate on nearly Keplerian orbits, highly inelastic
particle collisions can easily dissipate kinetic energy, and self-gravity of
granular material can be neglected in comparison to the gravitational
potential of the central object. Granular flows normally collapse into thin
disks, where particle number density increases and in some cases the flow can
be described using a fluid model with ``granular viscosity''.



It is known from the accretion disk theory that differentially rotating
viscous flows can be unstable \cite{LE1974,SS1976}. Indeed, it has been shown
that viscous instability sets in when the increase of surface density leads
to the decrease of the local viscosity \cite{LB1981,W1981,L1981}. In this
case, smallest density bump leads to the enhanced angular momentum transfer
and corresponding accretion process. Hence, mass accumulation at the outer
edge of the perturbation leads to further increase of density. Viscous
instability can operate in optically thick disks, where the viscous stress is
proportional to radiation pressure. However, phenomenological tests reveal
somewhat uncommon character of the instability, which even when occurred,
provides insignificant growth rates for linear  perturbations.


The second alternative energy source in Keplerian granular flows is the
viscous overstability \cite{K1978,BLY1984} that is thought to be a primary
mechanism for the development of some of the observed structures in dense
planetary rings. This axisymmetric pulsational instability occurs in granular
flows, where the derivative of kinematic
viscosity with respect to the surface density is positive and exceeds some
critical value \cite{W1981,PL1988,ST1995}. Thus, compressible epicyclic
response leads to viscous overcompensation and growth of density-spiral waves
due to increase of the viscous stress in the compressed phase. Later viscous
overstability has attracted considerable interest including its
non-axisymmetric \cite{B1985,LR1985,S2000,S2001,LO2006} as well as nonlinear
saturation properties \cite{ST1999,SS1999,LO2010}.

The key to the investigation of granular flows around gravitating objects is
a proper account for the particle collision effects. Kinetic description of
particle collisions has been successful in modelling properties of rapid and
dilute granular flows. Still, kinetic approach may fail due to the scale
separation problem between granular and flow time-scales and inelasticity of
particle collisions. In fact, it is known that a detailed theoretical
description of granular flows should deal with number of specific features:
granular gases are intrinsically non-equilibrium systems with non-Maxwellian
distribution functions that in some cases can reveal non-local of even
non-Markovian character (see Ref. \cite{Gold2003} and references therein).
Still, granular flows can be studied using hydrodynamic equations that can
describe collective phenomena including different types of instabilities,
thermal convection \cite{Khain2003,Pontuale2016}, behaviour of granular gas
mixtures, or clustering \cite{Olafsen1998,Brey1998,Gold1999,Serero2006}.

Significant advances in the understanding of the dense granular fluids have
been made recently. It seems that a wide range of dense granular flows can be
unified into a rheological model that permits formulation of a local
constitutive equation \cite{J2006,FP2008,BGP2011}. In this local rheological
model granular phenomenology is employed to define how fluid viscosity
depends on pressure, as well as strain tensor of the flow. Thus, granular
flow can be described by incompressible non-Newtonian fluid model, where
strain tensor is solely due to the velocity shear of the flow. We employ this
model for the description of astrophysical flows, where individual dust
granules can be highly porous particles colliding with a low restitution
parameters. In this limit dense granular flow can exhibits ``fluid''
properties even at moderate values of  particle volume fraction.

In the present paper, we study the linear stability of viscous Keplerian flow
around a gravitating center, taking into account rheological aspects of the
viscous stress tensor. Our incompressible model includes pressure and shear
rheology since they both affect linear stability of spiral waves. We identify
unstable axisymmetric modes analytically and analyze non-axisymmetric
instability numerically.

\textit{\textbf{Physical model.}}
The dynamics of an incompressible viscous flow rotating around central
gravitating object can be described by the Navier-Stokes equation:
\begin{equation}
\rho \left\{{\partial \over \partial t} + V_k {\partial \over \partial x_k}
\right\}
V_i = - {\partial P \over \partial x_i} + {\rho} {\partial \Phi \over \partial x_i} +
{\partial \tau_{ik} \over \partial x_k} ~,
\label{Eq_NS}
\end{equation}
where $\rho$, $P$ and $V_i$ are density, pressure and velocity of the flow,
respectively. We neglect self-gravity and assume that $\Phi$ is the
gravitational potential of the central object. The viscous stress tensor
$\tau_{ik}$ can  be calculated using the strain rate tensor
\begin{equation}
\tau_{ik} = \eta \dot \gamma_{ik} ~,~~~~
\dot \gamma_{ik} = {\partial V_k / \partial x_i} +
{\partial V_i / \partial x_k} ~,
\label{Tij}
\end{equation}
in incompressible  limit it is reduced to a shear strain tensor:
\begin{equation}
{\partial V_k / \partial x_k} = 0 ~.
\end{equation}
To describe the dissipative properties of the dense granular flow we employ
rheological fluid description implying the existence of a local constitutive
equation. Indeed, it has been shown recently, that granular fluids can be
described using the specific form of the non-Newtonian fluids (see Ref.
\cite{J2006} and references therein). In this limit viscosity of granular
fluid $\eta$ depends on both, pressure as well as the second invariant of the
strain rate tensor $\xi$:
\begin{equation}
\eta = \eta(P,\xi) ~,~~~
\xi =\sqrt{ \dot \gamma_{ik} \dot \gamma_{ik}  / 2 } ~.
\label{eta}
\end{equation}
This frictional visco-plastic constitutive law has been tested successfully
in laboratory experiments and is thought to be a general model describing
dense granular flows in ``fluid'' regime \cite{FP2008}. The ``fluid'' regime
of dense granular flows in laboratory is realized for a narrow range of
granular volume fraction, defined as the ratio of the volume occupied by the
grains to the total volume. Still, the granular rheology used here may also
work for lower density systems where the coefficient of restitution is low.

Alternative interpretation of the rheological model set by Eq. (\ref{eta})
can be obtained within the assumption of microscopic turbulence. Indeed,
Boussinesq eddy viscosity hypothesis assumes that turbulent viscosity
parameter can be calculated using the strain rate tensor (see Eq. 2). In such
limit, eddy viscosity can vary due to the variation of the intensity of
microscopic turbulence, depending on the pressure or local velocity shear of
the flow.


\textbf{\textit{Steady state.}}
Let us consider axisymmetric stationary differentially rotating viscous flow
in the cylindrical coordinates with constant pressure $\bar P$ and density
$\bar \rho$.
Azimuthal velocity of the background depends on the angular velocity of the
differential rotation $\bar V_{\phi} = r \Omega(r)$. The radial and azimuthal
components of the Navier-Stokes equation of the stationary state in polar
frame reads as:
\begin{eqnarray}
r \Omega^2 =
- {\partial \Phi \over \partial r} ~,
\label{Eq_NS_r}
\\
\left( r{\partial^2 \Omega \over \partial r^2}
+ 3 {\partial \Omega \over \partial r} \right) \bar \eta
+ r {\partial \Omega \over \partial r} {\partial \bar \eta \over \partial r}
= 0 ~,
\label{Eq_NS_phi}
\end{eqnarray}
where
\begin{equation}
\Phi(r,z) = {G M \over (r^2 + z^2)^{1/2}}
\label{Phi}
\end{equation}
is the gravitational potential of the central object with mass $M$. Assuming
thin disk model ($z^2/r^2 \ll 1$) we derive rotationally supported steady
state where the gravitational potential of central object sets Keplerian
profile of the angular velocity:
\begin{equation}
\Omega(r) = \Omega_0 \left( {r \over r_0} \right)^{-q} ~,~~~
{\Omega_0 = \left( {G M \over r_0^3} \right)^{1/2} ~.}
\label{Omega_K}
\end{equation}
Here $r_0$ is some fiducial radius used to parameterize the steady state and
$q=3/2$. Hence, using Keplerian angular velocity into the Eq.
(\ref{Eq_NS_phi}) we can derive radial profile of the viscosity parameter in
equilibrium:
\begin{equation}
{\partial \ln \bar \eta \over \partial \ln r } = q - 2 ~.
\label{q-2}
\end{equation}
Interestingly, Rayleigh stability criterion in rotating fluids $\partial_r
(r^2 \Omega(r)) >0$, or $q < 2$, indicates that in steady state, viscosity
parameter should be a decreasing function of radius: $ \partial_r \bar \eta <
0 $. Hence, Eqs. (\ref{Omega_K}) with radially homogeneous pressure and
density form the globally stable granular Keplerian flow that can be used for
the local linear stability analysis.

\textit{\textbf{Local linear analysis.}}
To study the linear dynamics of dense granular flows we split the velocity,
pressure and viscosity parameter into the background and perturbation
components:
\begin{eqnarray}
{\bf V} = {\bf \bar V} + {\bf V^\prime} ~,~~
P = \bar P + P^\prime ~,~~
\eta = \bar \eta + \eta^\prime ~.
\end{eqnarray}
We employ local shearing sheet approximation, where the flow curvature
effects can be neglected and the differential rotation is reduced to the
plane shear flow \cite{GL1964,T2003,T2010}. In this limit we expand azimuthal
velocity
\begin{equation}
\bar V_\phi(r) = r_0 \Omega_0 +
\left. {\partial (r \Omega) \over \partial r}\right|_{r_0} (r-r_0)
+ ...
\end{equation}
and use local approximation to neglect higher order terms with respect to
$(r-r_0)/r_0$. Hence, introducing the local Cartesian frame co-rotating with
the disk matter at the fiducial radius $r_0$
\begin{equation}
x = r - r_0 ~,~~ y = r_0 (\phi - \Omega_0 t) ~,
\end{equation}
and using standard form of the Oort constants
\begin{eqnarray}
A = {r_0 \over 2 } \left. {\partial \Omega \over \partial r}\right|_{r_0} ~,~~
B = -\Omega_0 - A ~,
\label{AB}
\end{eqnarray}
we can calculate steady state velocity
\begin{equation}
\bar V_y(x) = 2 A x ~
\end{equation}
that describes the radial shear of the azimuthal velocity due to the
differential rotation of the flow.

Hence, equation governing the linear dynamics of the
perturbations in local shearing sheet frame can be reduced to the following:
\begin{eqnarray}
{{\rm D~} \over {\rm D} t} V_x^\prime   - 2 \Omega_0 V_y^\prime
&=& - {1 \over \rho} {\partial P^\prime \over \partial x}
+ \nu \Delta V_x^\prime
+ {2 A \over \rho} {\partial \eta^\prime \over \partial y} ~,
\\
{{\rm D~} \over {\rm D} t} V_y^\prime - 2 B V_x^\prime
&=& - {1 \over \rho} {\partial P^\prime \over \partial y}
+ \nu \Delta V_y^\prime
+ {2 A \over \rho} {\partial \eta^\prime \over \partial x} ~,
\\
{{\rm D~} \over {\rm D} t} V_z^\prime
&=& - {1 \over \rho} {\partial P^\prime \over \partial z}
+ \nu \Delta V_z^\prime  ~.
\end{eqnarray}
where $\nu = \bar \eta / \rho$, ${\rm D}/{\rm D} t \equiv
{\partial / \partial t} + 2 A x {\partial / \partial y} $ and
$\Delta = \partial^2 / \partial x^2 + \partial^2 / \partial y^2 +
\partial^2 / \partial z^2$
and the radial gradient of viscosity parameter is neglected in the local
approximation: $\partial \bar \eta / \partial x = 0$.

To describe rheological properties of the flow we employ a general form of
the local constitutive equation and introduce pressure $G_P$ and shear $G_S$
rheology parameters as follows:
\begin{equation}
G_P \equiv \left({\partial \eta \over \partial P }\right)_{\xi} ~,~~
G_{S} \equiv {1 \over \rho} \left({\partial \eta \over \partial \xi }
\right)_{P} ~.
\label{GPS}
\end{equation}
Assuming that the rheological parameters of the granular fluid can be
considered to be locally constants we can calculate linear perturbation of
the viscosity as follows:
\begin{equation}
{\eta^\prime \over \rho} = G_{P} {P^\prime \over \rho} + G_{S}
\left({\partial V^\prime_{y} \over \partial x} + {\partial V^\prime_{x}
\over \partial y} \right) ~.
\end{equation}
Introducing Fourier expansion of the spatial variables in shearing sheet
frame
\begin{equation}
\left( \begin{array}{c}
{\bf V}^\prime({\bf r},t) \\
P^\prime({\bf r},t) / \rho \\
\eta^\prime({\bf r},t) / \rho
\end{array} \right)
\propto
\left( \begin{array}{c}
{\bf u}({\bf k},t) \\
- {\rm i} p({\bf k},t) \\
- {\rm i} \mu({\bf k},t)
\end{array} \right)
\exp \left({\rm i} {\bf r} {\bf k}(t) \right) ~,
\end{equation}
where ${\bf k}(t) = (k_x(t),k_y,k_z)$ and $k_x(t) = k_x(0)-2A k_y t$, we can
derive the system of equations governing the linear dynamics of
incompressible perturbations in time:
\begin{eqnarray}
\dot u_x(t) &=& 2 \Omega_0 u_y(t) - k_x(t) p(t) - \nu k^2(t) u_x(t) +
2 A k_y \mu(t) ~,
\nonumber \\
\dot u_y(t) &=& 2 B u_x(t) - k_y p(t) - \nu k^2(t) u_y(t) + 2 A k_x(t)
\mu(t) ~,
\nonumber \\
\dot u_z(t) &=& - k_z p(t) - \nu k^2(t) u_z(t) ~,
\label{ODE} \\
0 &=& k_x(t) u_x(t) + k_y u_y(t) + k_z u_z(t) ~,
\nonumber \\
\mu(t) &=& G_P p(t) - G_S (k_x(t) u_y(t) + k_y u_x(t) ) ~,
\nonumber
\end{eqnarray}
where $\dot \psi(t)$ stands for the time derivative of the variable $\psi(t)$
and $k^2(t) = k_x^2(t) + k_y^2 + k_z^2$. Equations (\ref{ODE}) pose a
complete initial value problem that can be solved numerically. However, to
get more insight into the stability properties of the system we derive an
approximate dispersion equation.

\textit{\textbf{Stability analysis.}}
Dispersion equation of the ODE system (\ref{ODE}) can be derived in the case
of rigid rotation ($A=0$). However, we employ adiabatic approximation when
time dependent mode frequency can be introduced and linear perturbations can
be expanded in time as: $ \psi(t) \propto \exp (-{\rm i} \omega(t) t) $. In
this limit we assume that frequency depends on time only through
the shearing variation of wave numbers: $\omega(t) = \omega({\bf k}(t))$.
Thus, the dispersion equation leads to:
\begin{equation}
\omega = \pm \left( \bar \kappa^2  - W^2 \right)^{1/2}
+ {\rm i} \left(W -\nu k^2 \right) ~,
\label{Disp}
\end{equation}
where $\bar \kappa$ sets epicyclic frequency in rheological flows:
\begin{equation}
\bar \kappa^2 = \left( - 4 B \Omega - 4 A^2 G_\gamma k_x k_y \right)
{k_z^2 \over k^2 - 4 A G_P k_x k_y} ~,
\end{equation}
and $W = \sigma_A + \sigma_P + \sigma_S$ with
\begin{eqnarray}
\sigma_A &=& {A k_x k_y \over k^2 - 4 A G_P k_x k_y } ~, \\
\sigma_P &=& 2 A G_P{(\Omega k_x^2 + B k_y^2) \over k^2 -
4 A G_P k_x k_y } ~, \\
\sigma_S &=& -A G_S{(k_x^2-k_y^2)^2 + k_\perp^2 k_z^2 \over k^2 -
4 A G_P k_x k_y } ~.
\end{eqnarray}
Here $\sigma_A$ describes the shear flow transient amplification due to the
differential rotation of the flow, while $\sigma_P$ and $\sigma_S$ describe
the effects of pressure and shear rheology, respectively.

In the rigidly rotating Newtonian fluids ($G_P=G_S=0$) solution reduces to
the classical spiral wave dumped by constant viscosity: $\omega = \pm 2
\Omega_0|k_z/k| - {\rm i} \nu k^2$.

The existence of growing modes can be seen in the case of differentially
rotating flows. Eq. (\ref{Disp}) shows that the necessary condition for the
growth of linear perturbations in differentially rotating granular fluids is
$W>0$. Therewith, the character of the perturbation growth depends on whether
rheological stress can destabilize epicyclic balance or not:
\begin{eqnarray}
\bar \kappa^2 > W^2 ,& W > \nu k^2  &: {\rm overstability}
\label{overstability}\\
\bar \kappa^2 < W^2 ,& W  + \sqrt{W^2 - \bar \kappa^2} > \nu k^2 &:
{\rm instability} \label{instability}
\end{eqnarray}

\textit{\textbf{Axisymmetric perturbations.}}
Eq. (\ref{Disp}) is rigorous in describing the stability of axisymmetric
modes with $k_y=0$. In this limit transient amplification is absent
($\sigma_A=0$), and we can analyze rheological modifications of the spiral
waves.

For the purpose of direct comparison with the viscous instabilities we
neglect shear rheology ($G_S=0$) and analyze the effect of pressure rheology
parameter. Then the necessary condition of the perturbation growth reduces
to:
\begin{equation}
G_P < 0 ~.
\end{equation}
This in turn indicates that the viscous overstability developing at $\partial
\eta / \partial \rho > 0$, i.e., $G_P>0$ is an intrinsically compressible
mechanism that is absent in the incompressible limit.

In the opposite limit, when pressure rheology can be neglected ($G_P=0$), we
recover new type of growth mechanism that originates from the shear rheology
of the granular fluid:
\begin{equation}
G_S > 0 ~.
\end{equation}
For better understanding we reformulate growth criteria
as $\sigma_S = -A G_S k_x^2 > \nu (k_x^2 + k_z^2)$. Hence, unstable modes are
nearly uniform in the vertical direction $|k_z/k_x| \ll 1$. Using Eqs.
(\ref{eta},\ref{AB},\ref{GPS}) and local value of incompressible
strain rate $\xi(r_0)=-2A$ we may rewrite the shear rheology instability
condition in a more general form:
\begin{equation}
\left( {\partial \ln \eta \over \partial \ln \xi} \right)_P > 2 ~.
\end{equation}
Thus, the shear rheology of the fluid leads to the visco-rotational
instability when the granular viscosity parameter increases faster than the
square of the shear (strain) rate.

In general, when pressure and shear rheology effects are comparable,
necessary condition of instability can be reduced to the following:
$\sigma_P + \sigma_S > \nu k^2 $.
Here we introduce the viscous cut-off wave-number $k_\nu$ that defines
length-scales that normally dissipate during one rotation period: $\Omega_0 =
\nu k_\nu^2 $. Hence, dynamically active modes are located in the $k/k_\nu <
1$ area of the spatial spectrum.

\begin{figure}[t]
\includegraphics[width=\columnwidth]{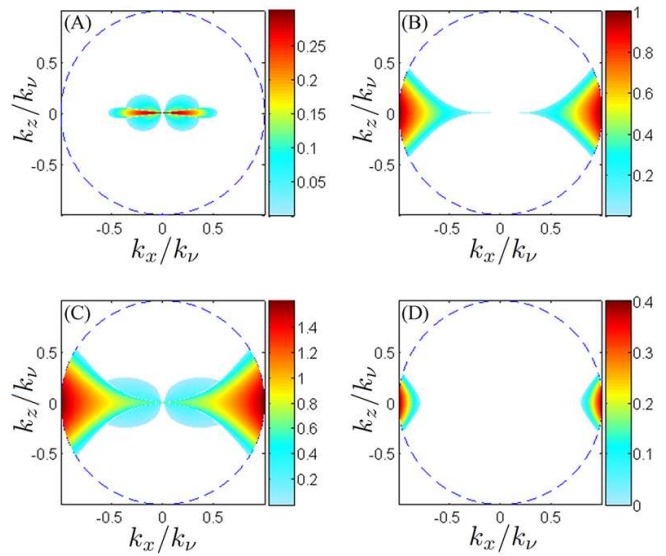}
\caption[]{(Color online)
Normalized growth rate of axisymmetric perturbations in granular fluids under
the influence of rheological viscous stress
${\rm Im}(\omega(k_x,k_z))/\Omega_0$ for different values of non-dimensional
pressure $g_p = \Omega_0 G_P$ and shear $g_s = \Omega_0 G_S / \nu$ rheology
parameters: (A) $g_P=-0.1, g_S=0$, (B) $g_P=0, g_S=1$,
(C) $g_P=-0.1, g_S=1$ and (D) $g_P=0.1, g_S=1$.
}\label{FIG_stability}
\end{figure}

The growth rates of linear axisymmetric perturbations are shown on Fig.
\ref{FIG_stability}. The growth mechanism due to pressure rheology favors
large-scale perturbations ($k_x/k_\nu \ll 1$, panel A), while shear rheology
instability operate at small radial scales ($k_x \sim k_\nu$, panel B). In
all cases most unstable modes are nearly uniform in the vertical direction
$k_z/k_\nu~\ll~1$. The growth rates of the visco-rotational instability set
by the shear rheology are asymptotically higher at wave-numbers larger than
the cut-off wave-number $k_\nu$. However, at length-scales shorter than the
granular dissipative scales the very validity of the rheological model breaks
down leading to the modification of the visco-rotational instability, a
process that we do not address in the current paper.

\begin{figure}[]
\includegraphics[width=\columnwidth]{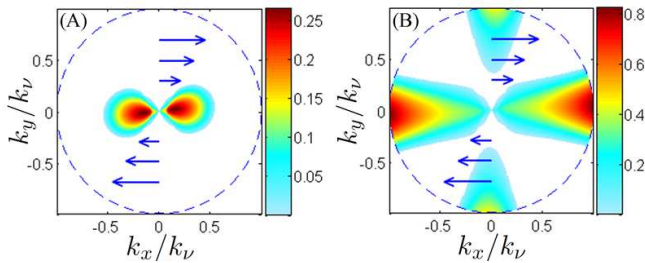}
\caption[]{(Color online)
Normalized growth rate of nonaxisymmetric perturbations in $k_x-k_y$ plane:
${\rm Im}(\omega(k_x,k_y))/\Omega_0$, $k_z/k_\nu=0.01$ and
(A) $g_P=-0.2, g_S=0$, (B) $g_P=-0.2, g_S=1.5$. Horizontal arrows indicate
wave-number drift due to the background shear.
}\label{FIG_nonaxisymmetric}
\end{figure}

\begin{figure}[t]
\includegraphics[width= 0.9 \columnwidth]{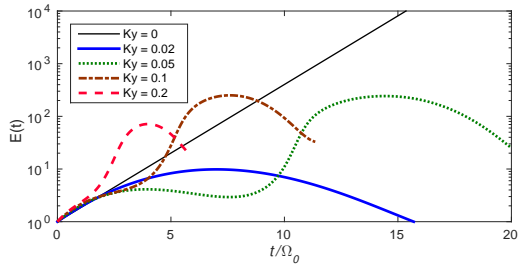}
\caption[]{(Color online)
Evolution of the energy of non-axisymmetric perturbations with
$g_p=0$, $g_s=1.5$, $k_x(0)/k_\nu=-0.8$, $k_z/k_\nu=0.01$ and different
values of azimuthal wave-number $k_y$. Modes with higher $k_y$ undergo
faster shearing deformation having less time to grow due to the
visco-rotational mechanism.
}\label{FIG_evolution}
\end{figure}

\textit{\textbf{Non-axisymmetric perturbations.}}
Linear dynamics of non-axisymmetric modes can be analyzed through Eqs.
(20-24), or numerical solution of the initial value problem (see Eqs.
\ref{ODE}). Fig. \ref{FIG_nonaxisymmetric} shows the growth rates in
($k_x,k_y$) plane. Shearing sheet modes are drifting in this plane due to the
background shear ($k_x=k_x(t)$). Thus, the non-axisymmetric modes have some
finite time before reaching viscous scale $k_\nu$, where they are dumped due
to a viscous dissipation. It seems that the pressure rheology parameter
introduces leading-trailing asymmetry of the linear modes: leading modes grow
higher for $G_P>0$, and trailing modes for $G_P<0$. Therewith, positive
pressure rheology decreases the growth rates of the shear rheology
instability, while the negative pressure rheology enhances it. Fig.
\ref{FIG_evolution} shows results of the numerical calculations of Eqs.
(\ref{ODE}). The energy of spiral waves is shown at different values of
azimuthal wave-number. Figure illustrates the transient character of the
growth of non-axisymmetric modes.

To get more insight into the nature of the instability
we derive dynamical equation in the limiting case of vertically uniform
perturbations ($k_z=0$) and shear rheology ($G_P=0$). We can
reformulate Eqs. (\ref{ODE}) for the horizontal velocity circulation:
\begin{equation}
{{\rm d} \over {\rm d} t} \left[ {\rm ln}
\left( {\rm curl}({\bf u})_z \right) \right]
= q G_S \Omega_0 {(k_x(t)^2 - k_y^2)^2 \over k(t)^2} - \nu k(t)^2 ~,
\label{curl}
\end{equation}
where ${\rm curl}({\bf u})_z = k_x(t) u_y - k_y u_x$ is the linear
perturbation of the horizontal vorticity and Oort's constant $A = -q \Omega_0
/ 2$. Hence we may conclude that visco-rotational shear instability of
horizontal vorticity perturbations occurs at $G_S>0$ in differentially
rotating flows with angular velocity decreasing outwards, and at $G_S<0$ if $q<0$.

\textit{\textbf{Summary.}}
We present the new type of instability in a rheological viscous dense
granular flows rotating around a central gravitating object. The
incompressible visco-rotational instability originates from the shear
rheology of the granular fluid. The instability operates on small scales and
differs in principle from the known viscous instabilities due to the pressure
rheology of viscous Keplerian flows. The mathematical formulation of the
problem is set to demonstrate fundamental nature of the found instability. We
adopt minimal model approach, showing that degrees of freedom necessary for
this instability to develop are 3 dimensionality and supercritical shear
rheology of the flow. The instability occurs in flows where the viscosity
parameter has a positive steep gradient with respect to the local shear
velocity. Unstable modes have small radial and large vertical scales,
indicating the possibility of instabilities for narrow azimuthal rings
(ribbons).

The visco-rotational shear instability can be simply described using the
pressure-vorticity balance. For instance, anticyclonic vorticity
perturbations to the Keplerian flow lead to local increase of the pressure.
When this vorticity increase leads to the increase of the viscosity and
corresponding accretion rate, pressure will increase even more, setting the
linearly runaway process. A similar process will occur with cyclonic
vorticity at pressure minima, for which a viscosity decrease will result in
further the flow pressure decrease.

The visco-rotational shear instability may lead to a nonlinear saturation at
higher amplitudes, or to the delocalization of the local constitutive
relation and deveopment of non-local structures due to the specific
properties of granular media \cite{KK2012}. We speculate that the instability
analyzed here can play a crucial role in the dynamics of dense planetary
rings, as well as promote structure formation in protoplanetary disks in the
areas of high dust to gas ratios.

\acknowledgements L.P. acknowledges support from the TSU Student Research
Council.

\end{document}